# An Electrical Grid with Discrete Energy Levels

H. Grebel

The Center for Energy Efficiency Resiliency Innovation (CEERI), NJIT, Newark, NJ 07102.
grebel@njit.edu

**Abstract:** Minimizing both power fluctuations and energy waste in an electrical grid is a central challenge to energy policy. Any discrepancy between power production and loads may lead to inefficiencies and instability in the system. Right now, the electrical grid is an analog system that only retroactively reacts to power demands. The balancing act becomes even harder with the penetration of sustainable resources (e.g., wind turbines). Here, we consider the effect of random perturbations to the grid's steady states operation. A model is constructed and analyzed within the framework of a randomly perturbed Markovian chain. Instead of balancing continuous values for supply and demand, the model assumes that both the generators and the users adhere to discrete pattern energy levels which is supplemented by local, short-term energy storage units (STESU). Under reasonable assumptions, we show that this grid maintains stability (meaning, a constant difference between supply and demand) over long periods of time while subjected to randomly fluctuating energy conditions.

## I. Introduction:

To maintain stability in presently deployed grids, power supply and power demand need to be balanced at all times [1]. Fluctuating demand and the penetration of intermittent sustainable sources make it a challenge to maintain grid stability, namely, the ability to maintain the power flow with no interruption, and power grid resiliency, namely, the ability to withstand disrupting events. Maintaining a constant level of generation and consumption power is hard to achieve in the present analog grid because the grid's response is reactive - the supply always follows the varying loads, and in principle, both may possess some degree of randomness. Approaches that use phase sensors [2] and predictive models have been implemented to make the grid controller aware of probable future instabilities [3]. These measures have difficulty to address demand, or supply fluctuations in real time. Grid stability and resiliency could potentially be achieved through energy-on-demand protocols, where energy requests are made ahead of delivery [4]. Such scenarios were simulated for a standard IEEE 39 bus for a limited channel capacity and searching for an optimal energy path [5]. Similarly to information networks, this approach requires loads' address and a high-speed communication system. .

In this manuscript, one concentrates on the generation and demand fluctuations. It attempts to build a grid that tolerates a certain amount of random fluctuations while maintaining a long term stability; another words, the grid attains a balance between energy generation and demand even if the average energy in the system fluctuates. Instead of balancing continuous power levels at all times, the grid delivers discrete energy units within a given time-frame. In order to accommodate this approach we need short term energy storage as buffers (or back up sources). The advantage of such approach as apparent below is that the grid may be balanced despite energy fluctuations in the system. While right now the grid is, basically, regulated by the nodes' impedance, a control system that translates the impedance changes into a discrete transition between discrete energy levels is needed.

Fluctuations are mathematically described here by random transition probabilities between pre-determined energy levels. Thus, introducing randomness to the transition probabilities represent a degree of system instability – this degree of instability would be analyzed well in advance in figuring out the best energy pattern that suits the grid's needs.

Overall, the paper attempts to find out if a balance between generation and demand can be reached even if the grid is randomly perturbed.

### I. a. Approach

Here, Energy terminology is used, rather than Power, because the power is assessed within some relatively long time-frame, longer than one second. The amount of deliverable energy adheres to pre-determined energy levels. An array of short-term energy storage units (STESU) helps maintaining the pre-determined energy pattern. While the average energy in the system (loads and generators) fluctuates, the energy storage units are charged and discharged in keeping with the pre-determined energy levels. The goal of the grid controller is to estimate the span of the energy random fluctuations ahead of time, and attempt to design an energy level system that mitigates the fluctuations' impact up to agreed-upon tolerance limits. Beyond these limits, the grid controller may use harsher methods, such as islanding. Mathematically, random behavior may be introduced to the grid by randomly perturbed Markov chains as a system model [6-9]. Markov chains are typically used to adapt the transition probabilities between states according to desired outcomes. Yet, for a demand system that is random, the outcomes are, in principle, unknown and the process of converging to a point of stability is unattainable. Instead of attempting to forecast the temporal grid behavior, it is suggested here to estimate its tolerance to instabilities. Rather than maintaining a global balance between generation and demand, which both have continuous values (the present analog approach), stability is maintained with pre-determined discrete energy values.

The balancing act of supply and demand may be relieved by using fast energy storage at the customer node (and to some extent at the generator) to regulate energy generation and consumption. In such scenario, energy may be delivered in packets with constant energy $U_0$. As a result, the energy levels of a grid node (known as a bus) are, $E_0=0$, $E_1$, $E_2$ and so on, which are multiples of $U_0$. For example, the energy $W_i$ that is delivered to node 'i', might be made of two units: $W_i=E_2-E_0=2U_0$. This is an extension of the concept of "power unit" [10] to "energy unit" which is delivered over longer time scales.

In power grids, the average power in the system fluctuates, the degree of which depends on the power grid's statistics. Yet, even small fluctuations are hard to keep up with because the generator needs to respond quickly (slow down or accelerate) to the fluctuating demands. Some of these fluctuations are self-healing but some may point to a deeper issues that could come to haunt the system later on. In an energy grid with discrete energy levels, both the load and the generator are forced into predetermined energy level pattern. The energy demand and the energy generation are still fluctuating, but the prediction of the energy requirement is limited to within this energy patterns. The STESU maintain the fluctuations at some percentage of this given level, say, $2\delta E/<E>=0.15$, with $<E>$, the average energy in the system and $\delta E$, the amplitude of the grid fluctuations. The response time of the STESU should be much faster compared to the time frame used to evaluate the energy - note that the present grid uses a fraction of a second as its time frame and a typical STESU has a much longer temporal response. The tolerance to uncertainties may be a learnt process - while we do not know the exact energy fluctuation at every time frame, we can determine its limits with some degree of confidence.

The simplest energy pattern is a 2-level system, with energy levels, $E_0$ and $E_1$ that could be called OFF and ON states. At this point, consider the case where states are powered by a stabilized generator. Further, assume that the transition from one time frame to another is Markovian with probability p denoting transition from level 0 to 1, and probability, q for a transition from 1 to 0. A typical Markovian process will settle to steady states levels after a short time with occupation vector of probabilities: $v_0=q/(p+q)$ and $v_1=p/(p+q)$ for each respective energy level. This, in turn will dictate the average energy in the system as $<E>=E_0 \cdot v_0+E_1 \cdot v_1$. The average energy value remains constant over time. One may call this system, the steady-states Markovian chain (*it is a steady-states rather than an equilibrium because one can prove that Markovian determinant is zero for every order of variation*).

Fluctuating demands make the transition probability 'fuzzy'; namely, $p \rightarrow <p>+\Delta p$ and $q \rightarrow <q>+\Delta q$ with $<p>$ and $<q>$, averaged values based on some long-term observations, and $\Delta p$, $\Delta q$ are random numbers that could be following a known statistics [9]. One may call it: the unstable grid. Obviously, the temporal behavior of the unstable grid is erratic but not completely out of control if the energy level are kept discrete - one may limit the grid operation close to those levels and invoke safety precautions upon deviations from it. In addition, the grid operator can no longer provide the grid with its average value since this value is fluctuating.

As implied by its name, the steady states Markovian chain reaches steady values after a short time, which is translated numerically to four to six time steps. For unstable cases, the average energy fluctuates because the initial conditions are randomly changing for each advancing time frame. In the case of a Gaussian distribution, the fluctuation amplitude is proportional to its width, $\sigma$, and its mean about which the average energy fluctuates.

<u>The power grid:</u> The current in a node 'i' that is part of a power network is affected by all nodes' voltages through,

$$I_i = \sum_j Y_{ij} V_j. \qquad j=1,2\ldots N \qquad (1)$$

Here $V_i$ is the node voltage and $Y_{ij}$ is the admittance matrix between node 'i' and node 'j'. The power grid is described by the power flow equations for the power flow, $S_i$ at the node 'i':

$$S_i = V_i \cdot [\sum_j (V_{ij} Y_j)]^* \equiv P + jQ \quad j=1,2\ldots N \qquad (2)$$

P and Q are the real (active) and imaginary (reactive) power components of the power flow $S_i$. The power flow is the transient of energy per second and Eqs. (1), (2) constitute a nonlinear set. All variable are continuous within some limits, set by boundary conditions (generator power, limits on the loads, etc.). The controller may take precautions, such as safety margins, and the generator is accelerated or slowed down based on the load.

<u>The energy grid:</u> Alternatively, time is divided into time-frames, $\Delta t$. The power is integrated over time,

$$W_i = \int_{t_1}^{t_2} dt' \, S_i, \qquad (3)$$

with $\Delta t = t_2 - t_1$. $W_i$ has dimensions of energy. One forces it to be a difference between two energy levels in the set $E_i$ - the possible energy levels for node 'i'. The simplest case would be that $W_i$ is composed of an integral number of an energy unit, $U_0$. In that case, $W_i = mU_0 = (E_i - E_j) \equiv \Delta E_{ij}$ with m - an integer number. Excess energy is absorbed and later released by the STESU. The controller provides energy to the load in pre-determined discrete format over the time-frame and uses the STESU as buffers.

**So how does it work?** The general energy level system is shown in Fig. 1. One may consider two main types: equally spaced (Fig. 1a) and clustered energy levels (Fig. 1b). The discrete energy level pattern and the statistic that governs its occupation state may be studied well in advance for various fluctuating scenarios.

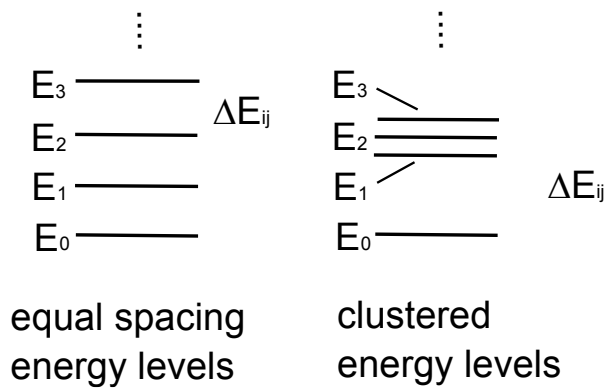

Figure 1. (a) Two types of an energy level system: equally spaced and clustered.

Energy pattern is constructed based on the long terms grid function (including the energy storage and an assessment of the grid fluctuations). Thus, the scenarios for most fluctuating possibilities have been worked out in advance. Right now, the relatively slow reaction of the grids to fluctuations lacks a reference point, which is provided by the pattern of the energy-levels.

For example, consider a two-state system discussed earlier. First, and as done now, calculate the power equations to each node, based on averaged historical data. Then, calculate the energy per time-frame per node (say, over 1 min) including the short-term energy storage capacity and a tolerance for fluctuation. For a two-level system, this value will be the energy provided by the generator, say $E=2\ U_0$ units. For a more elaborated system, changes in demand will force the generator to deliver an amount dictated by the next energy level, say $E=3\ U_0$ units. Energy may be provided in packets within a time frame [4]. A packet means a burst of energy (fixed absolute voltage, variable currents) for the duration of the time frame. For continuity, the packet does not have to refresh itself each time frame (known as non-return to zero scheme, NRZ) and may continue with different value of supplied voltage and current.

There are no energy fluctuation in "steady state" grid. Having a two level system in mind, the average energy in the system is smaller than the separation between the energy levels (because there is a finite probability that a node is down). The supplied energy to a particular node would be $E_1-E_0=mU_0$ for the grid to properly function. Interfacing "steady-state" grid with power storage allows the units to charge for a long-term back-up and for possible use in other applications. When the grid is made of many 2-level nodes, where some are ON and some are OFF, the supply will be closer to the average energy in the system. For a 2-level system, this may be represented as a sum,

$$E_{supplied}=\sum_i [p_i \cdot E^{(i)}_{node}+(1-p_i)\cdot E^{(i)}_{STESU}]\equiv mU_0 \qquad (4)$$

where, $p_i$ is the probability of a node to be ON, $E^{(i)}_{node}$ is the node energy for proper operation, and $E^{(i)}_{STESU}$ is the energy in the STESU both of which may be continuous. The supplied energy by the grid to the nodes is forced to be an integral of the energy unit, $U_0$.

The situation with an "unstable" grid is more complex even if the probability for the node to be down is small since the fluctuations drive the OFF state, as well. Again, and using a two level system as an example, fluctuations come from variable transitions between the OFF and ON states. For simulation purposes, the fluctuations are refreshed each time-frame. As a result, the average energy is fluctuating with some tolerance. As before, the grid controller allocates units of energy to each node – in the case simulated below, it is 2 energy unites per time-frame. In keeping up with the grid demands the STESU are charged and discharged. For simplicity, the dynamic of the STESU is not explicitly expressed here and it is part of the fluctuating transitions between the energy states. Namely, the statistics of the energy states implicitly includes them as loads, or local generators – by not including them we create highly unstable cases as seen from the graphs below. Under reasonable assumptions the system may become stable for many time frames even though the energy fluctuations are random.

The discrete energy pattern, its transitions and the fluctuation span may be a learnt process. Again, we do not know the exact nature of the fluctuations but we can put a limit to them. In analogy with the present power grid, the STESU play the role of safety margins with the added value that energy is not wasted but is used to charge them.

## II. Methods

In the following, a fluctuating grid is modeled as a perturbed Markov chain. Several scenarios are considered where, either the loads, or the generator, or both are fluctuating. The analysis involves fixed energy levels and clustered energy levels with various statistical transitions. Determining the transition probabilities is based on a learnt process; for example, the probability for the node (or the generator) to be down in a 2-level system might be very small compared to its normal operation, p>>q. At this point, the transmission line and inverters are included in the loads' characteristic.

As pointed earlier, the transition probability in an unstable grid, p, is p→<p>+Δp with <p>, averaged values based on some long-term observations, and Δp, a computer generated random numbers that are refreshed each time frame. A randomly sampled Gaussian distribution was chosen. Simulations of an unstable grid make a convenient use of random sampling, say, of the Gaussian distribution - it may not be limited to such distribution family and other distributions, such as scaled distributions should work equally well. The degree of such sampling process dictates the tolerance of the system to fluctuations [11]. Specifically, if one samples a Gaussian distribution by 10 points, it means that the confidence level of the probability for a fluctuation span lies within the 95% range of the distribution and the fluctuations take place across a width of an approximately four to five standard deviations, namely, $4\sigma$ to $5\sigma$. Sampling the Gaussian distribution by a 100 points means that the probability for the fluctuation span lies in the range of 99.5% of the distribution, or within ~$7\sigma$. Such numerical sampling method avoids cases where the probabilities exceeds the range of [0,1]. This method may be viewed as a predictive way for a fluctuation span range with a known confidence level.

The 'steady-states' cases are marked by blue lines whereas the unstable cases are marked by red lines. The random numbers used in the simulations followed a Gaussian distribution about a mean, with a variance, whose standard deviation is $\sigma$=0.02. The stability margin for a typical grid is ca 15%, or 2$\delta$E/<E>=0.15, with <E>, the average energy delivered and $\delta$E, the amplitude of the grid fluctuations. This particular choice of a standard deviation leads to a smaller change than a typical power grid margins: 2$\delta$E/<E>~0.2/1.9~0.11<0.15 (see for example, Fig. 2). In order to make fair comparisons, the chosen mean transition probabilities for the simulations favored the occupation of $E_1$=2, in units of energy, $U_0$. As we recall, the energy unit and its multiples are determined ahead of time based on the average loads' needs. The simulations employed a Markov chain. The 'steady-states' model was using the mean probability values and continued to use them for successive time steps. The unstable model used the mean probability in addition to a random number that perturbed it for every advancing time step – the random number was picked from the $5^{th}$ random sampling point of a Gaussian distribution that had a variance of $\sigma$=0.02.

## III. Results

### III.a. Fixed-level generator, fluctuating loads

Here we concentrate only on the loads because the generator provides a fixed amount of energy. We consider ensembles of equally separated 2-, 3- and 4-level and a clustered energy level systems.

For a 2-level system, $E_0=0$, $E_1=2$, $<p>=0.8$ and $<q>=0.051$. For a 3-level system, with $E_0=0$, $E_1=1$, $E_2=2$, all transitions probabilities leading to $E_2$ were taken as 0.8 and all other transition probabilities were taken as 0.051. Similarly, for a 4-level system, $E_0=0$, $E_1=1$, $E_2=2$ and $E_3=3$, all transitions leading to $E_2=2$ were taken as 0.8 and otherwise, as 0.051.

The average energy is defined as, $<E>=\sum_i E_i v_i$, where $E_i$ are the energy levels, $v_i$ is the steady state vector of probabilities for single packet occupancy and i is the energy level index, i=0,1, 2,...N. The span of energy fluctuation is well described by $2\delta E \sim 5\sigma <E>$ with reasonable results as we will see below.

<u>2-level system:</u> In the steady-states case, the system reached steady-states after 5 iterations (blue line in Fig. 2c) as a function of time steps. Fluctuations are noted for the unstable case, where the transition probability was randomly perturbed every time step. The fluctuations span was of the order of~ 0.25 units of energy for a mean average energy of $<E>\sim 1.88$ units of energy.

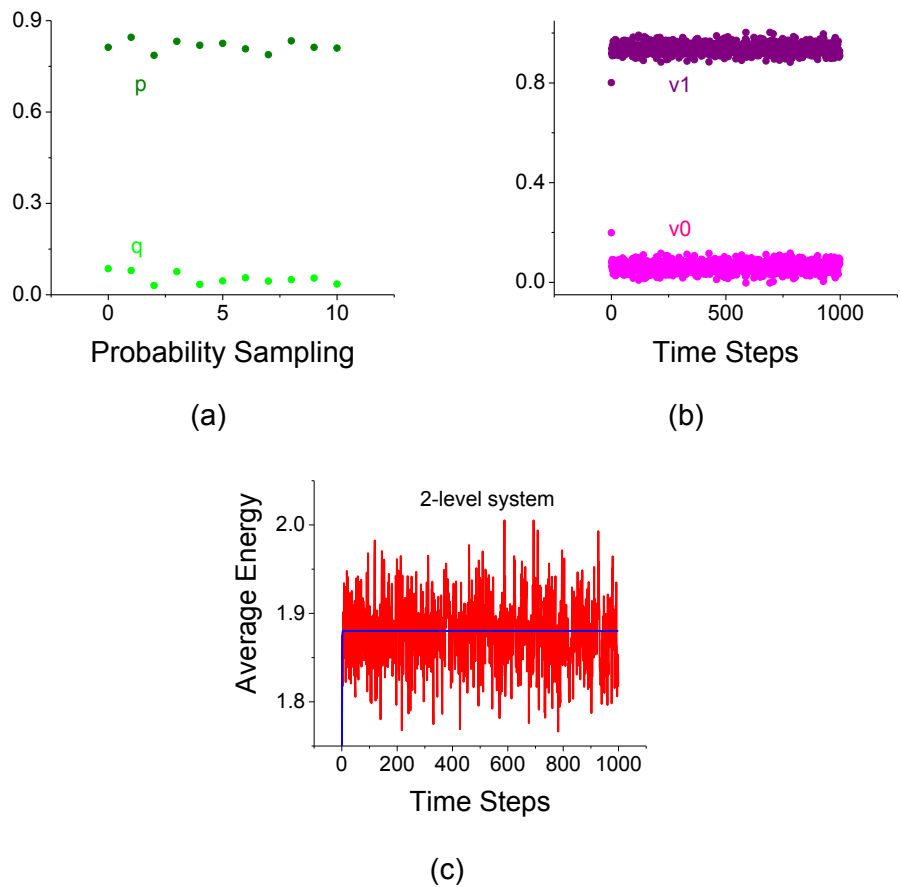

Figure 2. A 2-level system: (a) Sampling of the probability changes about $<p>=0.8$ and $<q>=0.051$ with 10 sampling points. In the simulation, the probability change was picked at the $5^{th}$ sampling point of a Gaussian distribution with $\sigma=0.02$. (b) Variations of occupation probability as a function of time for the unstable case. Pink: the OFF state; Dark purple: ON state. (c) Average energy for the steady-states case (blue) and the unstable case (red) as a function of time steps.

The relative average energy fluctuation was 2δE/<E>~0.25/1.85=0.13, similar to the fluctuation span in the occupation probabilities, ~0.13. In a 2-level system, the fluctuations in the E=2 level follows that of the average energy fluctuations since the OFF state carries no energy.

3- and 4-level systems: The systems maintained a steady-states after four iterations (blue line in Fig. 3). All transitions leading to the energy of choice (E=2 units) were taken with a probability of 0.8 whereas the probabilities of transitions that did not lead to that energy were taken as 0.051. The energy fluctuations span was of the order of~0.25 units of energy and a mean average energy of <E>~1.83 units of energy and 1.93 for the 3- and 4-level cases, respectively. The span of probability change was ~0.13. The relative average energy fluctuations, 2δE/<E> were similar to the variations in the occupation probabilities.

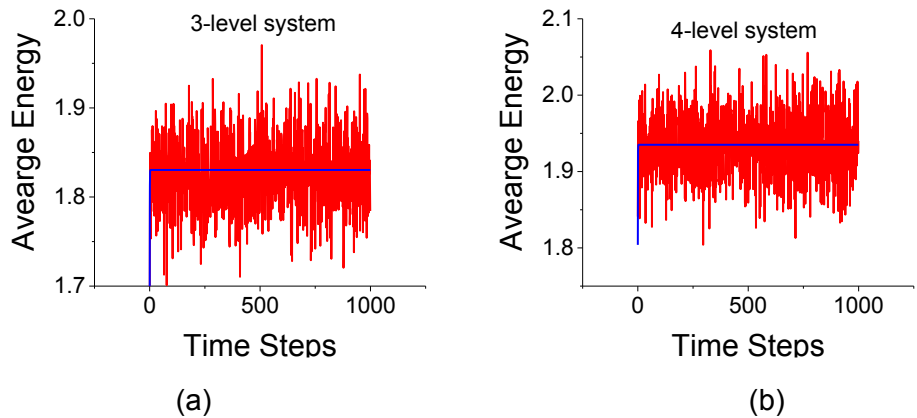

Figure 3. (a) A 3-level system, E:[0,1,2] and (b) a 4-level system, E:[0,1,2,3]: average energy for the steady-states case (blue) and the unstable case (red) as a function of time steps.

The effect of energy separation on the fluctuation span: The separation between the energy levels is related to node's demand. The transition probability is related to the uncertainty in this value. Thus, in assigning a transition probability we make a statement on a specific fluctuating scenario.

In Figs. 4, all transitions that were leading to the energy of choice of E=2 were taken with a probability of 0.8 whereas the probabilities of the other transitions were taken as 0.051. Fig. 4a shows the energy fluctuation span as a function of the standard deviation, σ. The dependence is linear as expected. Keeping the standard deviation fixed (either at σ=0.02 or at σ=0.025) and varying the separation between the energy levels exhibits a threshold at ΔE=0.3 beyond which the energy fluctuation span increases linearly. Thus, it means that a good strategy would be to keep the energy separation such that the fluctuation span is roughly constant. In assessing the average energy we used, $<E> = \sum_i E_i v_i$, where $E_i$ are the energy levels and the steady state vector of probabilities of occupancy for a single packet is $v_i$ and i=0,1,2,3.

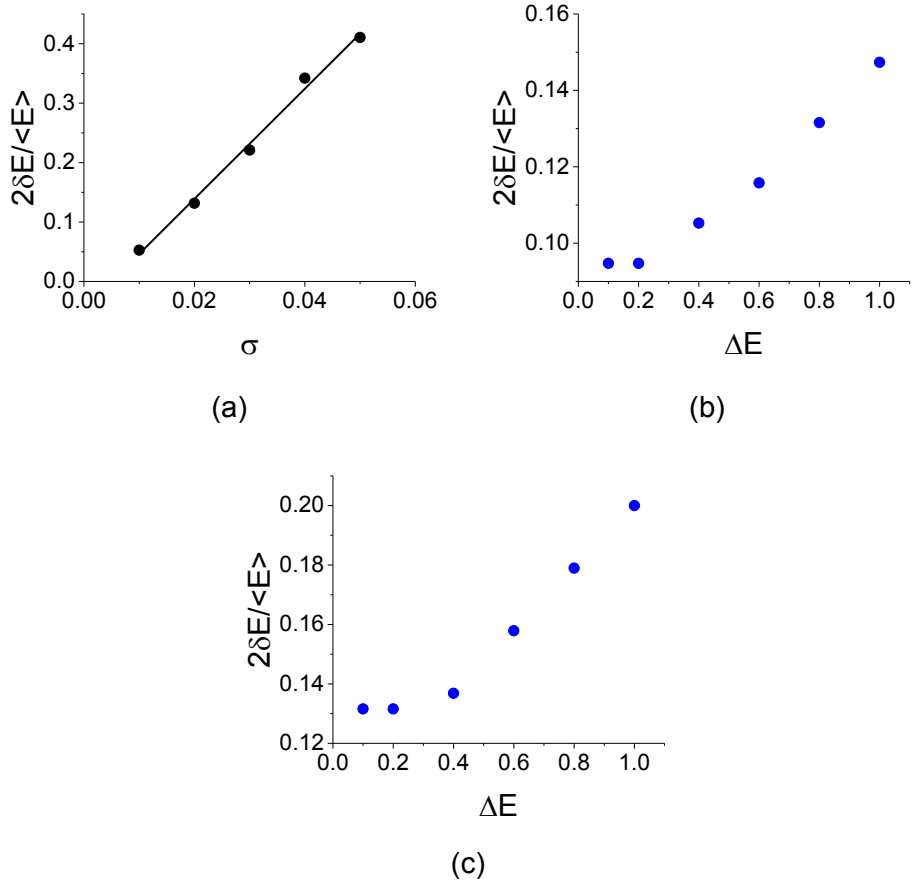

Figure 4. (a) Energy fluctuation span as a function of the standard deviation, σ, for E:[0,1,2,3]. Fixing the standard deviation at σ=0.02 (b), or at σ=0.025 (c), while varying the separation between the energy levels exhibits a threshold at ΔE=0.3 beyond which the energy fluctuation span increases linearly.

Clustered energy levels with a few preferential levels: Instead of a preferential transition to one particular energy level, the transition probabilities to any of the highest levels are made equal; in this case to, E:[1.8,2,2.2] from E=0. The separation between the higher energy levels was taken as 10% of the middle energy. This case may happen if the grid controller is not able to maintain a strict energy policy and the transition is spread over nearby levels. In Fig. 5a, the energy fluctuations are shown for two transition probabilities: 0.225 and 0.265 (note that the sum of probabilities along a Markovian matrix's column cannot exceed unity). The fluctuation span is similar for these two cases but smaller than the span for E:[0,1,2,3] of Fig. 4b.

**Locking to the highest probable energy level (the locked case):** In assessing the stability of the various fluctuating scenarios we gave credence to all possible occupation probabilities. Alternatively, the controller may opt to lock onto the highest probable energy level. For example, suppose the transition probabilities to the higher energy levels and amongst them is p=0.225. The occupation probabilities at time step 900 for E:[0,1.8,2,2.2] are, respectively, (0.073, 0.312, 0.315, 0.301). The controller picks the highest occupation probability (0.315 in this case) and forces the occupation probability vector to be (0,0,1,0), which means locking to energy request of

2 units. Such mathematical scenario is shown in Fig. 5b and its advantage is clearly seen as a zero fluctuation energy span. As the transition probabilities increases to 0.265, the fluctuation span increases but remains flat for long periods of time. Since the grid starts at an OFF state, E=0 with an occupation probability vector of (1,0,0,0) - the system will remain at the ground state when the transition probability is smaller than 0.22.

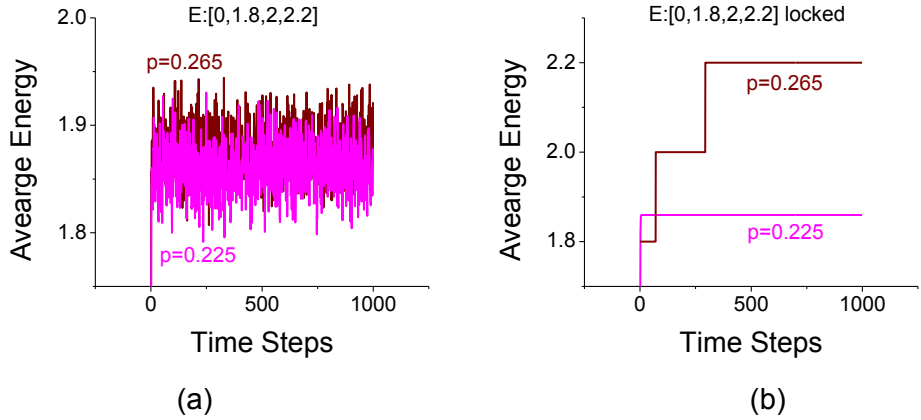

Figure 5. (a) The case where E:[0,1.8,2,2.2] and the transition is equally made to and between any of the higher 4 levels. The energy fluctuations are shown for two transition probabilities: 0.225 and 0.265. (b) The 'locked case: the controller identifies the highest occupation probability as the desired energy request and locks onto it. The grid is stabilized for longer periods of times as compared to (a).

### III.b. Fluctuating generator

Here, the generator is modeled as a 4-level system, $E_{gen}$:[0, 4.5, 5, 5.5] energy units. The energy gap between the clustered levels was kept at 10% of the central energy level, $E_3$=5 $U_0$. Such distribution allows for a zero power (e.g., in the case of wind turbine), albeit with a very low transition probability of 0.051. The transition to each of the higher levels from zero state was taken as 0.2 and so was the transition probabilities between them. The standard deviation was assumed as 0.025. Fig. 6a shows the average energy fluctuations over a time span of 1000 time frames. It indicates that the fluctuation span is ~$2\delta E/<E_{gen}>$~0.1~$4\sigma$, as expected. If one assumes that the generator is always on ON and that the lowest state (E=0) is eliminated (for example when using a back-up generator), then we are dealing with a 3-level generator as shown in Fig. 6b. In this case, $E_{gen}$:[4.5,5,5.5]. The fluctuation span is smaller in this case, $2\delta E/<E>$~0.02 when the transition probabilities are kept the same, 0.2 for proper comparisons.

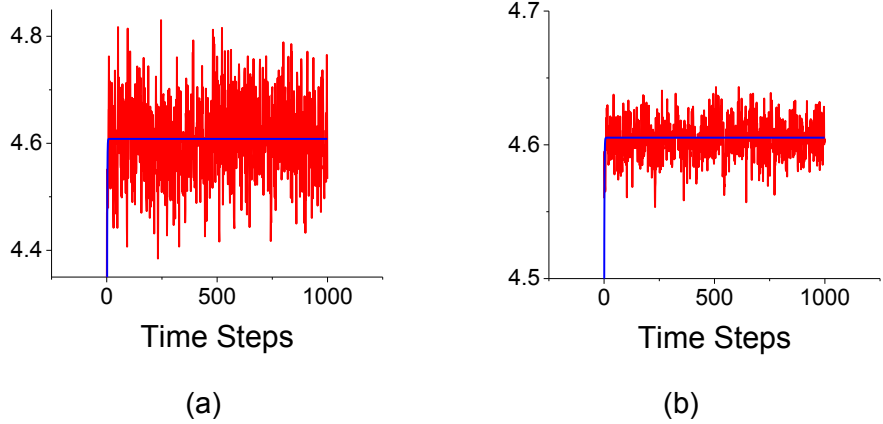

(a)                          (b)

Figure 6. (a) 4-level system where the energy levels are clustered around E=5. The transition preference is given to any of the high-energy levels, E:[0,4.5,5,5.5] at transition probability of 0.2. The transition to the OFF state is small, 0.051. (b) A 3-level generator, $E_{gen}$:[4.5,5,5] with transition probabilities of 0.2 among the levels.

### III.c. Fluctuating generator, multiple fluctuating loads

One may consider a generator with two loads. The generator is a 4-level system as depicted by Fig. 7a. The loads are chosen as, $E_{L1}$:[0,1.8,2,2.2] and $E_{L2}$:[0,2.7,3,3.3]; the generator has the energy levels as $E_{gen}$:[0,4.5,5,5.5]. The gaps are chosen as 10% of the central higher level. Fig. 7a shows the difference of energies (namely, the energy within the grid), $\langle\Delta\rangle=\langle E_{gen}\rangle-\sum_i\langle E_{Li}\rangle$ where i runs over all loads. As seen before, each individual load, $L_i$, exhibits a fluctuation span of the average energy, $2\delta E_{Li}/\langle E\rangle\sim 0.093$ or $\sim 4.6\times(\sigma=0.02)$. The transition probability was taken as 0.265.

If the loads are not correlated, then the average energy difference in the system would fluctuate about $\langle\Delta\rangle=0$. The fluctuation span of the difference would be the root of sum of all variances. Specifically, and in the case where all variance are similar, $N_L$ is the number of loads and $N_{gen}$ is the numbers of generators ($N_{gen}=1$ in our case),

$$\sigma_\Delta^2 = \sigma_{gen}^2 + \sum_i \sigma_{Li}^2 \sim \sigma^2(N_L+N_{gen}). \quad (5)$$

Thus, in the case of 2 loads and one generator, $\sigma_\Delta \sim \sigma\sqrt{3}$; with $\sigma=0.02$, one finds $2\delta\Delta/\langle E_{gen}\rangle \sim 4\sigma_\Delta \sim 0.14$, in a good agreement with the simulations of ca $2\delta\Delta/\langle E_{gen}\rangle \sim 0.13$.

If the generator fluctuations are fully correlated with the loads' fluctuations (as is the case for energy-on-demand scenarios), then the fluctuation span of the average energy difference, $2\delta\Delta/\langle E_{gen}\rangle$, would be zero.

The case where both the generator and each of the two loads are locked to the highest occupation probability is shown in Fig. 7b,c. Here, an equal transition probability is assumed between and to the higher energy levels. For a modest transitions' probability of 0.225 the energy difference in the system is constant, yet negative, meaning that the controller underestimated the energy needed and the difference will be provided by the energy storage units (up to some point). The energy fluctuation span is a bit larger for transition probabilities of 0.265 but exhibit long periods of flat response and the short-term energy units are mostly charged.

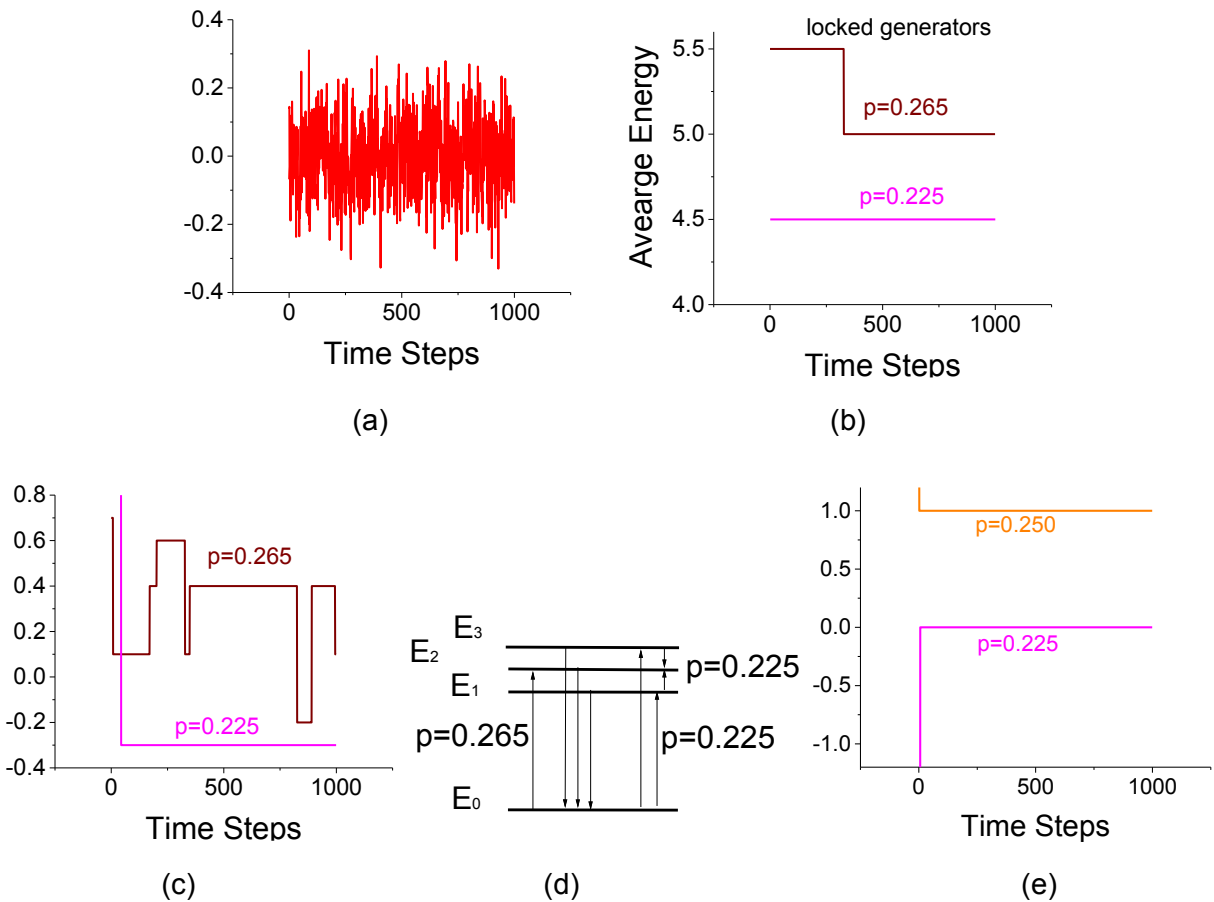

Figure 7. (a) Average energy difference for a 4-level unstable system, $<\Delta>=<E_{gen}>-\sum_i<E_{Li}>$. The fluctuation span, $2\delta E/<E_{gen}>\sim 0.13$, which is in good agreement with the analytical expression. (b) Simulation of a 4-level locked generator with two transition probabilities. (c) Average energy difference, $<\Delta>$, for a 4-level unstable system with a locked generator and two locked loads. The transition probabilities that lead to and in between the higher order energy levels are indicated. (d) Energy diagram with average transition probabilities. (e) Same as (c) with the exception that transitions leading to E=2 are with probability 0.265 while transitions between the higher order energy levels have a probability of 0.225, or 0.250.

One can rectify the long-term instabilities of Fig. 7c by further using knowledge of the transition probabilities themselves. In this case, the transition probabilities have preference to a specific energy level is shown in Fig. 7d. The transitions leading to E=2 have been accentuated. In essence, this case adds a distribution to the transitions to and from the preferred state as shown in Fig. 7d: the transitions leading to E=2 are with probability 0.265 while transitions between the higher order energy levels have a smaller probability of 0.225. The transition probability to the ground state (E=0) is small, p=0.051. As can be seen from Fig. 7e, the energy difference between the generator and loads is 0 most of the time. If the higher-order transitions are increased to

p=0.250, the energy difference remains constant at positive 1 unit. This means that the controller overestimated the energy requests, yet maintained the grid at a constant level. The extra energy will be charging the STESU, or an update to the energy level pattern is needed.

## IV. Discussions

An energy grid was examined with a statistical model that allowed the grid to fluctuate by a known tolerance, (e.g., $2\delta E/<E> \sim 15\%$; $2\delta E$ being the fluctuation span and $<E>$ being the average energy consumed by the loads). The energy in the grid assumes an array of energy levels and the fluctuations about these levels were assumed to be random. While the exact energy fluctuation at every moment is unknown, its limits may be determined with some degree of confidence. Beyond that tolerance limit (e.g., after a major disturbance, or a deliberate attack) the grid controller may resort to islanding. Alternatively, and having a 2-level system as an example, these undesired, blackouts scenarios may be simulated by a larger transition probabilities from the ON state $E_1$ level to the OFF state $E_0$. Larger values of fluctuation span may require a smaller energy level separation and the delivery of smaller energy units, $U_0$. In locking into the highest probable energy level not only simplifies the calculations but also stabilizes the grid for a long period of times. In practice, these simulations help the grid controller in figuring out the energy pattern and the tolerable fluctuation span (say, 15%). It also helps in the deployment of STESU and the duration of the time-frame.

Constructing the energy levels is related to the tolerance toward energy fluctuations. On one hand one would like to use as minimum number of energy levels as possible. On the other hand, one would like to map the load's needs as closely as possible without resorting back to the continuous (analog) scenario that is used today. A good strategy would be to keep the energy separation between levels such that the agreed upon fluctuation span is roughly constant (Fig. 4).

In assessing the effect of fluctuations, two major cases were considered: (1) the occupancy of each energy level and (2) digitizing (locking) the energy delivery to the energy with the highest occupancy value. The latter led to a long period of grid stability despite the randomness of the demand and supply.

The roll of energy storage units: it was implicitly assumed in previous scenarios, that the STESU efficiently buffer the loads when energy demands are randomly fluctuating. That means that the response of the energy storage units should be pretty fast on a time scale of a single frame – this is true for ordinary capacitors on time frames of seconds; super-capacitor with a response time of less than 10 second will be adequate for time frames of the orders of minutes [12-13]. Batteries would require longer charge-discharge times and hence longer time-frames. While right now, super-capacitors, or even batteries may not be as attractive due to energy densities and fast charging issues, combinations of this, in addition to fuel-cells (albeit with longer time-frames [14]) may be considered.

Advantages and disadvantages: The discrete energy level approach with STESU mitigate stability- and resiliency-risks. The STESU is a useful approach to energy that is otherwise wasted on safety margins and more research is needed to have reliable, high-voltage STESUs. On the other hand, for an impedance driven grid, the transition from one energy level to another may require high-voltage controllers [15].

## V. Conclusions

Under reasonable assumptions it was found that one may stabilizes the electric grid when delivering discrete units of energy (energy=power over a time-frame) instead of a continuous level of power (energy per second), even if the grid is randomly perturbed.

**Conflict of Interest Statement:** On behalf of all authors, the corresponding author states that there is no conflict of interest.

range. This does not mean that rare occasions may not happen – it only means that the focus is on those events that occur with some confidence level. Use of a finite sampling effectively is integrating over all, sometimes "improbable" cases within the sampling window.